\documentclass[conference]{IEEEtran}
\IEEEoverridecommandlockouts
\usepackage{cite}
\usepackage{amsmath,amssymb,amsfonts}
\usepackage{algorithmic}
\usepackage{graphicx}
\usepackage{textcomp}
\def\BibTeX{{\rm B\kern-.05em{\sc i\kern-.025em b}\kern-.08em
    T\kern-.1667em\lower.7ex\hbox{E}\kern-.125emX}}
\begin{document}

\title{Random Bit Generator Mechanism Based on Elliptic Curves and Secure Hash Function\\
\thanks{978-1-7281-1271-8/19/\$31.00 \textcopyright2020 IEEE.}
}

\author{\IEEEauthorblockN{Omar Reyad}
\IEEEauthorblockA{\textit{College of Computing and IT} \\
\textit{Shaqra University, Saudi Arabia}\\
\textit{Faculty of Science} \\
\textit{Sohag University, Egypt} \\
Email: ormak4@yahoo.com}
\and
\IEEEauthorblockN{Mohamed Karar}
\IEEEauthorblockA{\textit{College of Computing and IT} \\
\textit{Shaqra University}\\
Saudi Arabia \\
Email: mkarar@su.edu.sa}
\and
\IEEEauthorblockN{Kadry Hamed}
\IEEEauthorblockA{\textit{College of Computing and IT} \\
\textit{Shaqra University, Saudi Arabia}\\
\textit{Faculty of Computers and Information} \\
\textit{Minia University, Egypt} \\
Email: kadry@su.edu.sa}
}

\maketitle

\begin{abstract}
Pseudorandom bit generators (PRBG) can be designed to take the advantage of some hard number theoretic problems such as the discrete logarithm problem (DLP). Such type of generators will have good randomness and unpredictability properties as it is so difficult to find an easy solution to the regarding mathematical dilemma. Hash functions in turn play a remarkable role in many cryptographic tasks to achieve various security strengths. In this paper, a pseudorandom bit generator mechanism that is based mainly on the elliptic curve discrete logarithm problem (ECDLP) and hash derivation function is proposed. The cryptographic hash functions are used in consuming applications that require various security strengths. In a good hash function, finding whatever the input that can be mapped to any pre-specified output is considered computationally infeasible. The obtained pseudorandom bits are tested with NIST statistical tests and it also could fulfill the up-to-date standards. Moreover, a $256 \times 256$ grayscale images are encrypted with the obtained pseudorandom bits following by necessary analysis of the cipher images for security prove.
\end{abstract}


\section{Introduction} \label{Sec0}
A pseudorandom number generator (PRNG) which is also known as a deterministic random number generator (DRNG) is an algorithm for producing a string of random numbers whose characteristics could represent same properties of strings of random numbers produced from a truly random number generator (TRNG). The obtained number series from this PRNG is not considered truly random, because it is estimated by a comparatively set of known initial values, called the PRNG's \textbf{seed} \cite{c1}. Although PRNG strings are very close to TRNG sequences, thus could be generate by the use of hardware random number generators. Therefore, PRNGs are fundamental in real world applications for the speed in generate the numbers and also for their reproducibility. PRNGs are central in many recent applications such as electronic games, numerical simulations, statistical research, randomized algorithms, cryptography and lottery.

In the year 1985, elliptic curve public-key cryptosystems (also known as elliptic curve cryptography (ECC)), were introduced by Neal Koblitz \cite{c2} and Victor Miller \cite{c3}. By substituting the subgroup of the multiplicative group $\mathbb{Z}_{p}^{*}$ with the group of points on an elliptic curve (EC) over a finite field $\mathbb{F}$, these cryptosystems could be considered as EC analogues of the traditional discrete logarithm ones. The security beyond these EC cryptosystems is proven by the computational intractability support of the elliptic curve discrete logarithm problem (ECDLP). Since the ECDLP appears to be significantly harder than the ordinary discrete logarithm problem, the strength-for-key-bit is ultimately greater in elliptic curve based systems than in classical discrete logarithm based systems. Thus, parameters with small sizes is used in the case of ECC morethan with discrete logarithm systems with almost equivalent levels of security \cite{c4}. A worthy number of merits that could be acquired from the use of such small parameters include smaller key sizes, speed (faster computations) and secured certificates. Such features are important mostly in environments where bandwidth, storage space,  power of processing, or power consumption is bounded.

EC pseudorandom number generator (EC-PRNG) is considered to be the EC analogue of the previously mentioned PRNG. EC-PRNG is based mathematically on the ECDLP and is counted a cryptographically secure pseudorandom number generator (CS-PRNG). The idea of making use of ECs to be source of randomness dates back to early work introduced by Kaliski \cite{c5}. Since then, the use of elliptic curves for generating pseudorandom numbers has been
 studied for many years \cite{c6,c7,c8}. EC-PRNG logically can be separated into two main parts: one part that generates a sequence of EC points and one that extracts a bit string from that point sequence as a source of pseudorandom bits of uniform distribution.

In this study, we proposed a method for generating pseudorandom bit strings based on an elliptic curve points operations (Add, Double, Multiply) over finite fields ($\mathbb{F}_{p}$), which is secured by a hash function. The obtained pseudorandom bit string constructions are based on EC pseudorandom bit generator (EC-PRBG), which supported by a hash function to reinforce the security of an EC-PRNG. It is found that the statistical properties of the obtained bit strings can be greatly improved by applying this newly designed mechanism. Also, EC-PRBG mechanism is more suitable in producing pseudorandom bits in the EC-based digital signature algorithm (ECDSA) and session keys in related encryption stages.

This work is formulated as next. The preliminaries of EC are introduced in Section \ref{Sec1}. An overview of elliptic curves over $\mathbb{F}_{p}$ are mentioned in Section \ref{Sec1.1}. In Section \ref{Sec1.2}, the related EC point operations are presented. The cryptographic hash function is discussed in Section \ref{Sec1.3}. In Section \ref{Sec2}, the pseudorandom bit generator mechanism is proposed. Experimental and NIST test results are presented in Section \ref{Sec3}. In Section \ref{Sec4}, an image encryption  application with various security analysis is given and finally conclusions are given in Section \ref{Sec5}.

\section{Preliminaries} \label{Sec1}

\subsection{Elliptic Curves over $\mathbb{F}_{p}$} \label{Sec1.1}

Lets consider an EC to be $E$ over $\mathbb{F}_{p}$, $p>3$, given by the following affine \textit{Weierstrass} equation in the form 

\begin{equation} \label{EQ1.1}
E: y^{2} =x^{3} + ax + b,  
\end{equation}

\noindent where $a$ and $b$ are coefficients belonging to $\mathbb{F}_{p}$ such that $4a^{3} +27b^{2} \ne 0$. The set of $\mathbb{F}_{p}$ consists of all the points $(x,y), x \in \mathbb{F}_{p}, y \in \mathbb{F}_{p}$, which fulfill the defining equation (\ref{EQ1.1}), together with a special
point $O$ called the point at infinity \cite{c2}.

\textbf{Example 1.} Let a prime $p = 29$ and consider the elliptic curve $E: y^{2} =x^{3} + x + 4$ defined over $\mathbb{F}_{29}$. This curve has order $33$ and is cyclic. Note that parameters values are $a = 1$ and $b = 4$ and $4a^{3} +27b^{2} = 4+ 432 = 436 \equiv 1 \: (\text{mod} \, 29)$, in this case $E$ is considered as an EC. Also points of $\mathbb{F}_{29}$ are point $O$ and the other points which listed in Table \ref{tab3.1}.

\begin{table}
\caption{Elliptic Curve Points in $\mathbb{F}_{29}$}
\label{tab3.1}       % Give a unique label
\begin{center} 
\begin{tabular}{l|l|l|l|l|l|l|l}
\hline\noalign{\smallskip}
$(0, 2)$   &  $(0, 27)$  &  $(1, 8)$   &  $(1, 21)$   &  $(3, 11)$  &  $(3, 18)$    \\             
$(6, 9)$   &  $(6, 20)$  &  $(7, 8)$   &  $(7, 21)$   &  $(10, 12)$ &  $(10, 17)$   \\
$(12, 2)$  &  $(12, 27)$ &  $(14, 6)$  &  $(14, 23)$  &  $(15, 1)$  &  $(15, 28)$   \\
$(17, 2)$  &  $(17, 27)$ &  $(18, 5)$  &  $(18, 24)$  &  $(19, 3)$  &  $(19, 26)$  \\
$(20, 7)$  &  $(20, 22)$ &  $(21, 8)$  &  $(21, 21)$  &  $(25, 9)$  &  $(25, 20)$  \\
$(27, 9)$  &  $(27, 20)$  \\  
\noalign{\smallskip}\hline
\end{tabular}
\end{center} 
\end{table}

\subsection{Elliptic Curve Point Operations} \label{Sec1.2}
\begin{figure} [h!]
\centering
\includegraphics[width= 0.43 \textwidth]{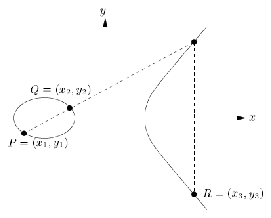}
\caption{Description of two EC-points addition: $P + Q = R$}
\label{Fig121}       
\end{figure}

The addition operation of two points on an elliptic curve $E(\mathbb{F}_{p})$ to result in a third point on same curve, the chord-and-tangent rule is used. With this addition operation, the set of points $E(\mathbb{F}_{p})$ forms a group with $O$ serving as its identity. The formed group is used in the elliptic curve cryptosystem structure. Let $P=(x_{1} ,y_{1})$ and $Q=(x_{2} ,y_{2})$ to be two featured points on an elliptic curve $E$. The sum of such points $P$ and $Q$, denoted $P + Q = R =(x_{3} ,y_{3})$, is defined as next. First, a line is drawn through points $P$ and $Q$; this line intersects the same EC in a third point as drawn in Figure \ref{Fig121}. As a result, point $R$ is the reflection of that process in the $x$-axis \cite{c9}. 

Consider point $P=(x_{1} ,y_{1})$, the double of $P$ which is denoted $2P = R =(x_{3} ,y_{3})$, is obtained as follows. The tangent line with EC at point $P$ is drawn first. This line intersects the EC in a second point. As a result, point $R$ is the reflection of this process in the $x$-axis as shown in Figure \ref{Fig122}. The sum of two points, also the double of one point is deduced from algebraic description as follows:

\begin{figure} [h!]
\centering
\includegraphics[width= 0.43 \textwidth]{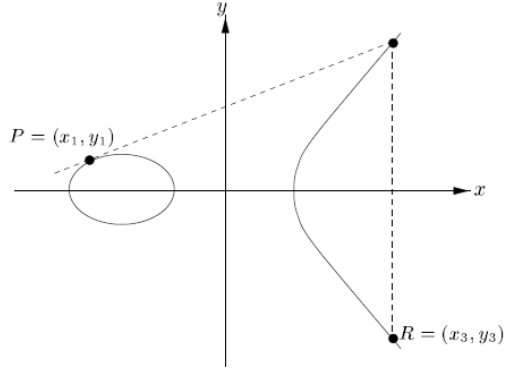}
\caption{Description of one EC-point doubling: $P + P = 2P = R$}
\label{Fig122}       
\end{figure}

\begin{enumerate}
\item Note that $P + O = O + P$ \quad all of $P \in E(\mathbb{F}_{p})$. 
\item For point $P = (x, y) \in E(\mathbb{F}_{p})$, so that $(x, y)+ (x, -y) = O$, the point $(x, -y)$ is indicate as $-P$, so it is called the negative of $P$; notice that $-P$ is in fact a point on the same curve $E$. 
\item Addition of points $P= (x_{1}, y_{1}) \in E(\mathbb{F}_{p})$ and $Q= (x_{2}, y_{2}) \in E(\mathbb{F}_{p})$, with $P \neq \pm Q$. The operation $P + Q = R = (x_{3} ,y_{3})$, where 

\begin{equation} \label{EQ1.2}
\begin{array}{c c c}x_{3} &=\left(\frac{y_{2} -y_{1} }{x_{2} -x_{1} } \right)^{2} -x_{1} -x_{2}   \\
\text{and} \\
y_{3} &=\left(\frac{y_{2} -y_{1} }{x_{2} -x_{1} } \right)(x_{1} -x_{3} )-y_{1} \\ \end{array} .
\end{equation}

\item Doubling of point $P= (x_{1}, y_{1}) \in E(\mathbb{F}_{p})$, with $P \neq -P$. The operation $P + P = 2P = (x_{3} ,y_{3})$, where 

\begin{equation} \label{EQ1.3}
\begin{array}{c c c} x_{3} &=\left(\frac{3x_{1}^{2} +a}{2y_{1} } \right)^{2} -2x_{1} \\
 \text{and} \\
 y_{3} &=\left(\frac{3x_{1}^{2} +a}{2y_{1} } \right)(x_{1} -x_{3} )-y_{1} \\  \end{array} .
\end{equation}
 
\end{enumerate}

Note that, the addition of these two points $P$ and $Q$ in $E(\mathbb{F}_{p})$ needs some arithmetic operations such as (addition, subtraction, multiplication, and inversion) in the $\mathbb{F}_{p}$ field .

\textbf{Example 2.} Let EC defined in Example $1$ used again.
\begin{enumerate}
\item Points $P = (3, 11)$ and $Q = (14, 23)$, so that $P + Q = (x_{3}, y_{3})$ can be computed as: 

\begin{equation} \label{EQ1.4}
\begin{array}{c c c} x_{3} =\left(\frac{23 - 11}{14 - 3} \right)^{2} - 3- 14 = 64 \equiv 6 (\text{mod} \; 29) \end{array} 
\end{equation}

\noindent and 

\begin{equation} \label{EQ1.5}
\begin{array}{c c c} y_{3} = 9(3- 6)- 11= -38 \equiv 20 (\text{mod} \; 29) \end{array} .
\end{equation}

Resulting $P+ Q= (6, 20)$

\item Point $P = (3, 11)$ which doubled as $2P = P+ P =(x_{3}, y_{3})$ can be computed as: 

\begin{equation} \label{EQ1.6}
\begin{array}{c c c}x_{3} =\left(\frac{3(3^2) + 1}{22} \right)^{2} - 6 = 619 \equiv 10 (\text{mod} \; 29) \end{array}  
\end{equation}

\noindent and

\begin{equation} \label{EQ1.7}
\begin{array}{c c c} y_{3} = 25(3- 10)- 11= -186 \equiv 17 (\text{mod} \; 29) \end{array} .
\end{equation}

Hence, $2P= (10, 17)$.
\end{enumerate}

\subsection{Secure Hash Function} \label{Sec1.3}
The PRBG mechanism can be based on a non-invertible or one-way hash function \cite{c1}. The hash-based EC-PRBG mechanism which used here is designed to make use of any suitable secure hash which is used by exhaustion applications that need different security strengths, providing that a suitable hash function is utilized and adequate entropy is gained for the seed value.

Thereafter, $H$ hash function is defined as:

\begin{equation} \label{EQ1.3.1}
H_i = f(X_i, H_{i-1}) \quad i = 1,2, \cdots, t
\end{equation}

From the equation, $f$ be the round function, $H_t$ is the hashcode and $H_0$ is equal to an initial value (IV). For hash function safely usage ought to concur or exceed the in demand security strength of the exhaustion applications. 

\begin{figure} [h!]
\centering
\includegraphics[width= 0.48 \textwidth]{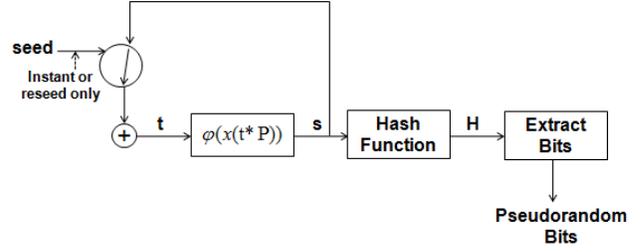}
\caption{Pseudorandom Bit Generator Mechanism}
\label{Fig1N}       
\end{figure}

\section{The Proposed Random Bit Generator Mechanism} \label{Sec2}
The proposed EC-PRBG is based on the hardness of ECDLP which can be described as: given two points $P$ and $Q$ on EC of order $n$, how to get $a$ such that $Q = aP$?. The EC-PRBG mechanism is depicted in Figure \ref{Fig1N}.

The instantiation stage of the EC-PRBG mechanism requires choosing an appropriate EC and points on that curve for the required security strength. The seed value which is used to locate the initial value $(s)$ of the EC-PRBG must have enough bits of entropy with sufficient security strength. The value of $t$ is accounted for the seedlen-bit number in the initial state, so we can consider that $t = s_{0}$ in this case. The EC-PRBG can offer security strength as long as the security strength of the used curve. The main reason of using the hash function $H$ is to ensure that the entropy is distributed throughout the extracted bits, provided that they are verifiably random.

\begin{figure} [h!]
\centering
\includegraphics[width= 0.26 \textwidth]{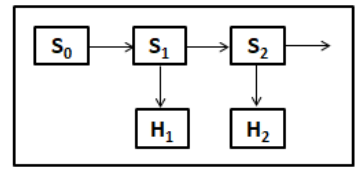}
\caption{EC-PRBG Backtracking Resistance}
\label{Fig2N}       
\end{figure}

Backtracking resistance in this mechanism is deep-seated, even in the case that the internal state is vulnerable to exposure. As shown in Figure \ref{Fig2N}, EC-PRBG generates a seed value for each step $i = 1 ,2 ,3 ,\cdots, $ as follows:

\begin{align*} \label{EQ2.1}
s_{i} &= \varphi (x([s_{i-1}])P) \quad , i=1,2,..., \\ 
H_{i} &= \varphi (x(s_{i}),H_{i-1}) \quad , i=1,2,...,  
\end{align*}

\noindent where $s_{0} \in E(\mathbb{F}_{p})$ is the "initial value". The EC-PRBG method represents an EC scalar multiplication with the extraction of the $x-$coordinate from the resulting points $s_{i}$ and from the random hash output $H_{i}$ with truncation operation to obtain the output pseudorandom bits. Following a line in the same direction of the arrow is the normal operation; inverting that direction reveals the ability to solve the ECDLP for that specific curve. The ability of an adversary to invert the arrow in Figure \ref{Fig2N}, implies that the adversary has solved the ECDLP for that specific elliptic curve. Backtracking resistance is built into the mechanism design, as knowledge of $s_1$ does not allow an adversary to determine $s_0$ (and so forth) unless the adversary is able to solve the ECDLP for that specific curve. Furthermore, knowledge of $H_1$ does not allow an adversary to determine $s_1$ (and so forth) unless the adversary is able to solve the ECDLP for that specific curve.

The EC-PRBG generates pseudorandom bit strings by extracting bits from an EC points. The internal state of the EC-PRBG is a secret value $s$ that represents the $x$-coordinate of a point on an EC. Output bits are produced by first computing $H$ to be the $x$-coordinate of the point $[s]P$, and then extracting low order bits from the $x$-coordinate of the hashcode output $H$.

\section{Experimental Results} \label{Sec3}
The implementation of the presented EC-PRBG mechanism need to include an approved curve. Once the designer chooses the security level  required by a given application, he can then start the implementation of an EC that most NIST SP 800-90A \cite{c1} appropriately meets this requirement. 

\subsection{Implementation Example} \label{Sec3.1}
The EC-PRBG algorithm allows an exhaustion application to instantiate using a prime curve. In accordance security key strengths of 112, 128, 192 and 256 bits may then be supported. The secure hash algorithm (SHA)-256 is chosen as the hash function. SHA-256 function generates an almost-unique, fixed size 256-bit hash. In this experiment, the implementation process used the following EC equation:

 \begin{equation} \label{EQ3.1.1} 
 E: y^{2} =x^{3} + 4 x + 1               
 \end{equation}

 \noindent where $a = 4, b = 1$ are $E$ parameters over $\mathbb{F}_{p}, p = 503$ and the cardinality of $E$ is $N = \#E(\mathbb{F}_{503}) = 516$. Also the generator point $G = (283, 315)$ of order $\ell = 129$ is selected for our EC-PRBG mechanism.

\subsection{The NIST Randomness Tests} \label{Sec3.2}
The NIST \cite{c11} test suite is a statistical package that consisting of up to $15$ tests. It is developed for testing the randomness of binary strings obtained by either hardware or software based cryptographic pseudorandom and random number generators. The $15$ tests concentrate on a variety of different non-randomness types that could exist in a bit strings. The proposed mechanism produces a very random bit strings as reflected by the high p-values as shown in Table \ref{NIST}.

%{\begin{table} [!ht]
{\begin{table} [htbp]
\caption{Test results for 1048576 bit strings}
\label{NIST}
\begin{center}
\begin{tabular}{| l | c | c |} \hline
\textbf{Test-name}        & \textbf{P-value} & \textbf{Result} \\ \hline
Block Frequency (m = 100)        & 0.046169  & Succeed     \\ \hline
Frequency                        & 0.681211  & Succeed     \\ \hline
Cusum (Forward)                    & 0.878529  & Succeed     \\ \hline
Cusum (Reverse)                    & 0.674391  & Succeed    \\ \hline
Long Runs of Ones                & 0.128851  & Succeed     \\ \hline
Spectral DFT                     & 0.149590  & Succeed     \\ \hline
Rank                             & 0.638151  & Succeed     \\ \hline
Lempel Ziv Complexity            & 1.000000  & Succeed      \\ \hline
Overlapping Templates (m = 9)    & 0.120402  & Succeed     \\ \hline
NonOverlapping Templates (m = 9) & 0.197506  & Succeed \\ \hline
Approximate Entropy (m = 10)     & 0.681211  & Succeed     \\ \hline
Universal (L = 7, Q = 1280)      & 0.051599  & Succeed     \\ \hline
Random Excursions (x = +1)       & 0.297235  & Succeed     \\ \hline
Serial (m = 16)                  & 0.343750  & Succeed     \\ \hline
Random Excursions Variant (x = +1)& 0.050388 & Succeed     \\ \hline
Runs                             & 0.499889  & Succeed     \\ \hline
Linear Complexity (M = 500)      & 0.703017  & Succeed    \\ \hline
\end{tabular}
\end{center}
\end{table}}

\section{Application In Image Encryption} \label{Sec4}
\begin{figure} [h!]
\centering
\includegraphics[width= 0.45 \textwidth]{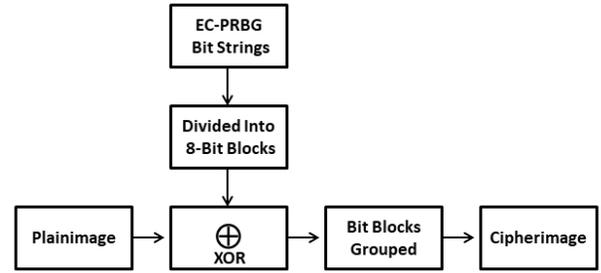}
\caption{EC-PRBG Based Image Encryption Schema}
\label{Figmain}       
\end{figure}

With the huge amount of the data and high correlation between the adjacent pixels in images, stream ciphers are highly preferred over block ciphers in image encryption applications. The security of digital images need pseudorandom bit strings that have pretty good randomness properties and also high periodicity. Recently, several works that using ECs for digital and medical images encryption has been presented in literature such as \cite{c12,c13,c14,c14a,c15}. In this work, the EC-PRBG is used for encrypting a $256 \times 256$ grayscale of Lena image pixels as shown in Figure \ref{Figmain}. Each image pixel has a $8$-bit value of between $0$ and $255$, so the pseudorandom bit strings in turn divided into blocks of $8$-bit each. Next, bitwise XOR operation is carried on every bit of the $8$-bit block. The resulted bits then grouped together to obtain the cipherimage. The decryption process is done vice-versa and the following security analysis is carried out.

\begin{figure} [h!]
\centering
\includegraphics[width= 0.44 \textwidth]{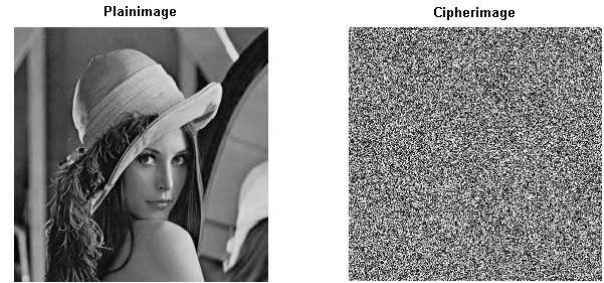}
\caption{Lena plainimage and the corresponding cipherimage with the EC-PRBG mechanism}
\label{Fig3}       % Give a unique label
\end{figure}

\subsection{Entropy Analysis} \label{Sec4.1}
The entropy $H(m)$ of a message source $m$ is calculated from the equation:

\begin{equation}
\label{EQ4.1.1}
H{(m)} = - \sum\limits_{k=0}^{255} {P{(m_{k})}log_{2}}{P(m_{k})}
\end{equation}

\noindent where $P(m_{k})$ represents the probability of message $m_{k}$ \cite{c16}. The various entropy values for Lena plain and encrypted image which shown in Figure \ref{Fig3} are indicated in Table \ref{tab3.2}. It is remarkable that the entropy of the encrypted image is too close to the theoretical value of $8$ which elucidate that all of the pixels in the encrypted image occur with almost equal probability. Therefore, the proposed EC-PRBG is secure against the entropy-based attack and the information leakage is negligible. 

{\begin{table}[h!]
\caption{Entropy and basic parameters for Lena image}
\label{tab3.2}       % Give a unique label
\begin{center}
\begin{tabular}{ l | c | l | l | l } \hline
Scheme           & Entropy & PSNR     & MAE    & MSE      \\  \hline
Proposed EC-PRBG & 7.9971  & 8.5631   & 79.52  & 9305.32  \\  \hline
Ref\cite{c18}    & 7.9898  & 8.5838   & -----   & 9009.33  \\  \hline
Ref\cite{c19}    & 7.9968  & 11.30    & 79.22  & 4859.03  \\
\end{tabular}
\end{center}
\end{table}}

\subsection{Mean Absolute Error and Mean Square Error} \label{Sec4.2}
The cipherimage must demonstrates a significant difference with it's corresponding plainimage. This difference can be measured by two major techniques, Mean Absolute Error (MAE) and Mean Square Error (MSE) \cite{c17}. MAE and MSE values are calculated by using the following equations:

\begin{equation}
\label{EQ4.2.1}
MAE = \frac{1}{W*H} \sum\limits_{j=1}^{H} {\sum\limits_{i=1}^{W} {|(P_{ij}-C_{ij})|}}
\end{equation}

\begin{equation}
\label{EQ4.2.2}
MSE = \frac{1}{W*H} \sum\limits_{j=1}^{H} {\sum\limits_{i=1}^{W} {(P_{ij}-C_{ij})^2}}
\end{equation}

In equations (\ref{EQ4.2.1}) and (\ref{EQ4.2.2}), parameters $W$ and $H$ are the width and height of that image. Also $P_{ij}$ is the gray level of the pixel in the plainimage and $C_{i,j}$ is the gray level of the pixel in the cipherimage. MAE and MSE values of the cipherimage are reported in Table \ref{tab3.2}. As shown from the table, MAE and MSE tests have produced  high values which then guarantee the resistance of the EC-PRBG mechanism against differential attacks.

\subsection{Peak Signal-to-noise Ratio (PSNR)} \label{Sec4.3}
PSNR is mainly used in image processing area as a consistent image quality metric \cite{c20} and the greater PSNR, the better the output image quality. The performance of the proposed EC-PRBG method is estimated on the basis of PSNR and the measure values obtained are shown in Table \ref{tab3.2}. The obtained results clearly illustrated that the EC-PRBG mechanism is well suited for many types of image encryption operations.

\subsection{Correlation Analysis} \label{Sec4.4}
For any common image, two neighboring pixels in a plainimage are strongly correlated vertically, horizontally and diagonally. The maximum value of correlation coefficient is $1$ and the minimum value is $0$ \cite{c21}. Horizontal, vertical and diagonal directions results are obtained as shown in Table \ref{tab3.3} for plainimage of Lena and for it's ciphered image by the EC-PRBG method respectively. The obtained results elucidate that there is negligible correlation between the two adjacent pixels in the cipherimage, even when this two adjacent pixels in the plainimage are highly correlated as shown in Figure \ref{Fig5}.

\subsection{Sensitivity Analysis} \label{Sec4.5}
If one small change in a plainimage able to cause a significant change in the corresponding cipherimage, with respect to diffusion and confusion properties, then the known-plaintext attack actually loses its efficiency and becomes practically useless. To quantify that demand, two joint measures are used: Number of Pixels Change Rate (NPCR) and Unified Average Changing Intensity (UACI) \cite{c22}. The test results shown that the average values of the percentage of pixels changed in cipherimage is greater than 99.47\% for NPCR and 30.48\% for UACI for the pseudorandom bits. This means that the EC-PRBG method works perfectly and precisely with respect to small changes in the plainimage pixels.

\begin{figure} [h!]
\centering
\includegraphics[width= 0.50 \textwidth]{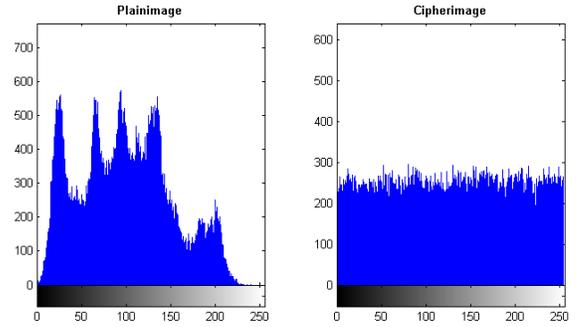}
\caption{Histogram of Lena cipherimage with the EC-PRBG operation}
\label{Fig4}       % Give a unique label
\end{figure}

\subsection{Histogram Analysis} \label{Sec4.6}
To block the information leak to an adversary, an image encryption schema should always produce a cipherimage of the uniform histogram for all of the corresponding plainimage \cite{c17}. The histograms for Lena plainimage and cipherimage are estimated. Lena plainimage histogram contains large spikes while the histogram of it's cipherimage is almost flat and uniform as depicted in Figure \ref{Fig4} which denotes equal probability of occurrence of each pixel. Histogram of Lena cipherimage is remarkably different from the respective plainimage and consequently does not provide any evidence to appoint known statistical attacks on the image encryption application. 

\begin{figure} [h!]
\centering
\includegraphics[width= 0.53 \textwidth]{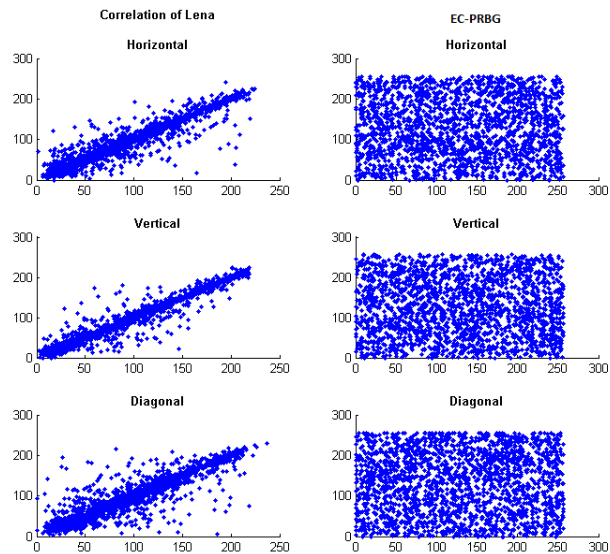}
\caption{Correlation of Lena plainimage and cipherimage with the EC-PRBG}
\label{Fig5}       % Give a unique label
\end{figure}
%=====================================================

{\begin{table}[h!]
\caption{Correlation coefficients for Lena image}
\label{tab3.3}       % Give a unique label
\begin{center}
\begin{tabular}{ l | l | l | l } \hline
Scheme           & Horizontal & Vertical   & Diagonal  \\ \hline
Plainimage       & 0.93915    & 0.96890    & 0.91686   \\ \hline
Proposed EC-PRBG & -0.00287   & 0.03450    & 0.00196   \\ \hline
Ref\cite{c18}    & -0.00041   & -0.00025   & -0.000027   \\ \hline
Ref\cite{c19}    & -0.0043    & -0.0090    & -0.0031    \\
\end{tabular}
\end{center}
\end{table}}

\section{Conclusion} \label{Sec5}
This paper presented a new mechanism for generating pseudorandom bit strings based on elliptic curve group over finite fields ($\mathbb{F}_{p}$) and hash derivation function to achieve high security levels. The performance analysis and security results showed that the obtained bit strings have high periodicity and good randomness properties. Moreover, an application in image encryption based on cipher bit strings stream was examined and various security analysis of the cipherimage is reported.

\section*{Acknowledgment}
The prepared work has been supported financially by Shaqra University, Saudi Arabia and Sohag University, Egypt.

%\section*{Acknowledgment}

%\section*{References}

\end{document}